\documentclass{aastex631}

\usepackage{CJK}

\shorttitle{Flare Location}
\shortauthors{Li et al.}

\begin{document}

\begin{CJK*}{UTF8}{gbsn}

\title{Locate a stellar flare from the M dwarf LAMOST J1332+5057}

\author[0000-0001-7515-6307]{Guang-Wei Li (李广伟)}
\affiliation{Key laboratory of Space Astronomy and Technology, National Astronomical Observatories, Chinese Academy of Sciences, Beijing, 100101, China}

\author[0000-0001-8228-565X]{Gui-Ping Zhou (周桂萍)}
\affiliation{State Key Laboratory of Solar Activity and Space Weather, National Astronomical Observatories, Chinese Academy of Sciences, Beijing, 100101, China}

\author[0000-0001-6820-6441]{Bing Du (杜冰)}
\affiliation{Key Laboratory of Optical Astronomy, National Astronomical Observatories, Chinese Academy of Sciences, Beijing, 100101, China}

\author[0000-0002-7640-5368]{Yan-Xin Guo (郭炎鑫)}
\affiliation{Key Laboratory of Optical Astronomy, National Astronomical Observatories, Chinese Academy of Sciences, Beijing, 100101, China}

\author[0000-0002-9909-3758]{Wei-Hua Wang (王伟华)}
\affiliation{School of Computer Science and Information Engineering, Changzhou Institute of Technology, Changzhou, 213032, China}



\begin{abstract}

Young M-type stars exhibit frequent flares, which would seriously impact their habitable planets. Since stellar surfaces cannot be resolved, flare locations remain unknown. Here, by using the Mg I b emission line in LAMOST spectra, the location of a stellar flare from a young M dwarf is pinpointed at $ (-123.0^{+8.0}_{-5.8}, 80.5^{+2.9}_{-3.2})$ in degree in the polar region. 
Our method can be used to accurately locate stellar flares. This would enable us to assess the impact of stellar flares on planets more accurately and improve our understanding of stellar dynamo models.
\end{abstract}



\section{Introduction} \label{sec:intro}

Solar flares primarily occur at low latitudes ($<30^\circ$) \citep{2018AdSpR..62.2701A}. Flares accompanied by coronal mass ejections (CMEs) pose risks to technological infrastructure on Earth \citep{2015AdSpR..55.2745S}. 
Young stars tend to frequently produce energetic flares \citep{2021A&A...645A..42I,2019ApJ...871..241D}. Even M-type stars exhibit vigorous flaring activities with energies of up to $10^{35}$ erg \citep{2019ApJS..241...29Y,2024ApJ...971..114L}.
 Direct impact of these energetic stellar flares on exoplanets would result in extreme space weather \citep{2019AsBio..19...64T,2007AsBio...7..185L}.
 \par
 However, some observational evidence suggests that stellar flares can occur in high-latitude regions.
\citet{2021NatAs...5..697V} reported a coronal dimming following a flare from the star AB Dor. The data show no obvious rotational modulation and the duration of dimming is longer than the stellar rotational period. Based on these facts, it was hypothesized that the dimming was caused by a CME released by a flare located in the polar region. By modelling flare profiles and assuming flare region were  circular, \citet{2021MNRAS.507.1723I} located four high-latitude flares on different stars, while \citet{2024A&A...682A.176B} located three high-latitude flares on the star CD-36 3202. Conversely, by comparing stellar flares with solar flares at different latitudes, \citet{2025A&A...695A..21Y}  demonstrated that stellar flares primarily occur around equators.
\par
If CMEs released by high-latitude flares travelled in a straight line, they would only have a marginal affect on planets located in the ecliptic plane. However, these high-latitude CMEs may be deflected. For M dwarfs with strong magnetic fields, smaller CMEs can become trapped in current sheets, whereas larger CMEs are deflected only slightly \citep{2016ApJ...826..195K}. For Sun-like stars, \citet{2019ApJ...886L..37K} found that the magnetic field of the young star $k^1$ Cetii deflects its CMEs, increasing their impact on its planets by 30\%. \citet{2023MNRAS.522.4392M} studied CMEs from stars Kepler-63 and Kepler-411, finding that higher-velocity CMEs exhibit smaller deflections, while strong magnetic fields lead to significant deflections. By comparison, solar CMEs always exhibit small deflections. In summary, although CMEs occurring at high latitudes may affect planets, their overall impact on planets will be reduced compared to CMEs occurring near the equator. Therefore, to accurately assess the impact of CMEs on habitable planets, their locations should be determined precisely.
\par
Observations suggest that spots tend to appear in high-latitude regions on rapidly rotating stars \citep{2009A&ARv..17..251S}. One possible explanation is that rapid rotation causes magnetic flux tubes tend to emerge in these regions \citep{1992A&A...264L..13S}. Additionally, \citet{2022ApJ...928...51B} demonstrated that rapidly rotating stars may also exhibit longitudinal patterns at low latitudes. Therefore, if the locations of active regions could be determined by flares, the dynamo mechanism would be better understood.
\par
In the flaring spectra of LAMOST  (Large Sky Area Multi-Object Fiber Spectroscopic Telescope, also known as the Guoshoujing Telescope) medium-resolution spectra (MRS; $R = \frac{\lambda}{\Delta \lambda} = 7500$)  \citep{2020arXiv200507210L}, the Balmer  H$\alpha$ line is the most prominent. As the wavelength is given in vacuum, the wavelength of H$\alpha$ is 6564.61 ${\rm \AA}$ in LAMOST spectra. However, as it is present throughout the entire active region, including the chromosphere, corona, arcades, ribbons, flux ropes, and magnetic loops, its velocity components are highly complex. The chromospheric Mg I b emission lines from M dwarfs are also very strong in the LAMOST flaring spectra \citep{2022A&A...663A.140L}, and their widths are significantly narrower than the H$\alpha$ line. Furthermore, on the Sun, Mg I lines generally only exist in the coolest chromospheric regions \citep{2017A&A...604A..50S}. Therefore, in this work, we will use the Mg Ib line to trace the location of a flare. In fact, the ionization energy of Mg I is $\sim 7.65$ eV, which is significantly lower than the upper energy level of H$\alpha$ ($\sim$ 12.1 eV). This also indicates that the region producing the Mg I b line has a lower temperature than the region producing H$\alpha$. 

%
%

\section{Flare Spectra}
We searched all MRS in LAMOST DR10 by detecting the variations of H$\alpha$ emission lines, and found a white-light flare from the young active M-type star LAMOST J133213.33+505701.8  (hereafter J1332+5057), on April 19, 2022. The observation (\textsf{obsid = 1007903112}) of the flare lasted for 2.89 hours, and was recorded continuously by eight individual spectra at a 22-minute cadence, which are called flare spectra. On May 8, 2022,  LAMOST observed this star again (\textsf{obsid = 1011803112}) and the coadded spectrum was used as the quiescence spectrum. All LAMOST spectra are not flux-calibrated, and the spectral fluxes are given in counts. 
\par
A flare spectrum, $f(\lambda)$, is the sum of the stellar quiescence spectrum  $f_{\rm Q}(\lambda)$ and the flare-only spectrum $f_{\rm flare}(\lambda)$ which is produced by the flare. The $f_{\rm flare}(\lambda)$ consists of the continuum (denoted as $f_{\rm flare, con}(\lambda)$) and the emission lines (denoted as $f_{\rm flare, em}(\lambda)$) spectra.  That is:
\begin{eqnarray} \label{equ:qf}
f(\lambda) &=& f_{\rm Q}(\lambda)  + f_{\rm flare}(\lambda) \\
                 &=& f_{\rm Q}(\lambda)  + f_{\rm flare, con}(\lambda) + f_{\rm flare, em}(\lambda)
\end{eqnarray}
The quiescence spectrum of \textsf{obsid = 1011803112} is denoted as $f_{\rm Q, 0}(\lambda)$, and different from the quiescence spectrum in the flare spectrum $f_{\rm Q}(\lambda)$ because these two spectra were obtained in different times, but $f_{\rm Q}(\lambda) = a f_{\rm Q, 0}(\lambda)$, where $a$ is an unknown constant. \par
We assume that the flare continuum can be described as a parabola (that is, in Equation~\ref{equ:qf}, $f_{\rm flare, con}(\lambda) =  b \lambda^2 + c\lambda +d$, where $b$, $c$ and $d$ are constants), so the flare spectrum without emission lines can be written as: 

\begin{equation} \label{equ:fs}
f(\lambda) - f_{\rm flare, em}(\lambda)  =a f_{\rm Q, 0}(\lambda)  + b \lambda^2 + c\lambda +d
\end{equation}

\begin{figure}[ht]
  \centering
   \includegraphics[scale=0.5]{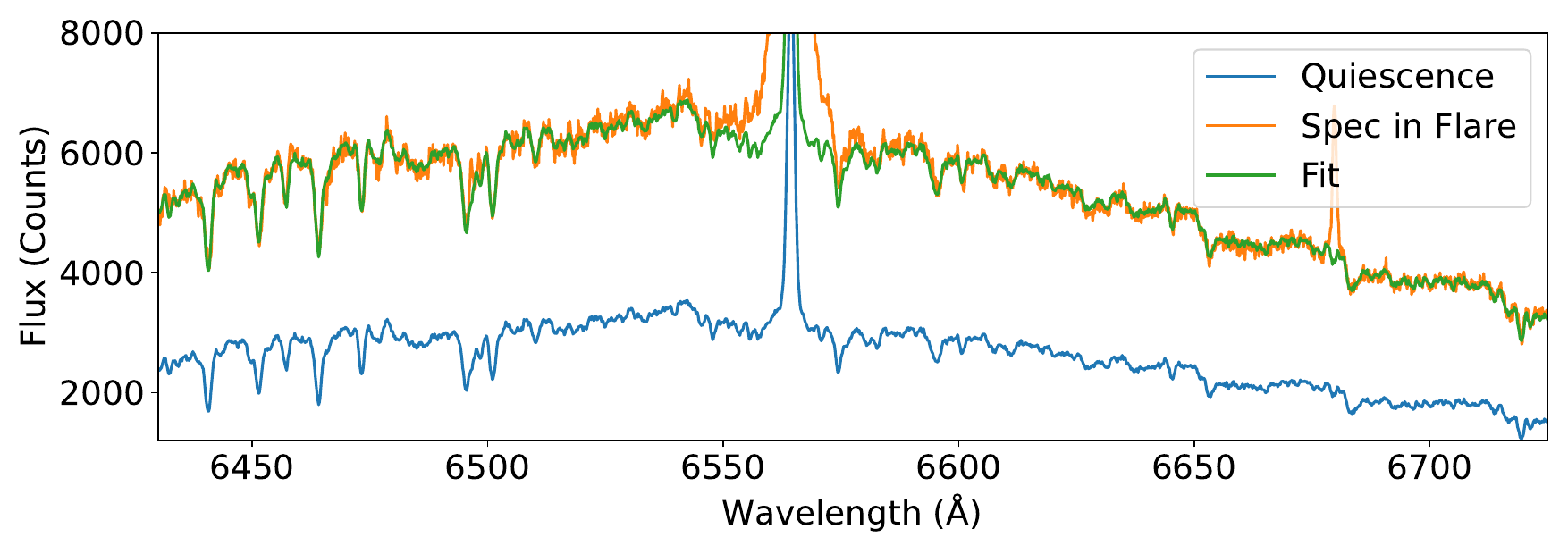}
  \caption{The quiescence spectrum and a spectrum during flare. The quiescence spectrum, the first spectrum during flare and the fitted spectrum by Equation~\ref{equ:fs} are shown in blue, brown and green, respectively.} 
\label{fig:emspec}
\end{figure}

In 6430 \AA $< \lambda <$ 6725 \AA, there are only two emission lines: H$\alpha$ and He I $\lambda$6680 in a flare spectrum (see Fig.~\ref{fig:emspec}). We removed these two emission lines and fitted each flare spectrum using Equation~\ref{equ:fs} to obtain $a$, $b$, $c$ and $d$. 
In Fig.~\ref{fig:emspec}, the quiescence spectrum of \textsf{obsid = 1011803112} ($f_{\rm Q, 0}(\lambda)$), the first flare spectrum ($f(\lambda)$) and the fitted spectrum by Equation~\ref{equ:fs} ($a f_{\rm Q, 0}(\lambda)  + b \lambda^2 + c\lambda +d$) are shown in blue, brown and green, respectively.

\par

Now, the flare-only $f_{\rm flare}(\lambda)$ is:

\begin{equation} \label{equ:fc-cali}
f_{\rm flare}(\lambda) = \frac{1}{a}f(\lambda) - f_{{\rm Q},0} (\lambda)
\end{equation}

and the emission spectrum $f_{\rm flare, em}(\lambda)$ :

\begin{equation} \label{equ:em-cali}
f_{\rm em}(\lambda) =f_{\rm flare}(\lambda) - \frac{b \lambda^2 + c\lambda +d }{a }.
\end{equation}
where, 6430   \AA $< \lambda < $6725  \AA.
\par

Since a small variation from each $f (\lambda_i)$ or $f_{{\rm Q},0} (\lambda_i)$, has almost no effect on $a$, $b$, $c$, and $d$ in Equation \ref{equ:fc-cali} and \ref{equ:em-cali}, we neglected the covariance between these parameters ($a$, $b$, $c$, and $d$) and $f (\lambda_i)$ or $f_{{\rm Q},0} (\lambda_i)$. Therefore, for a pixel $i$, the error $\sigma_i $ of $f_{\rm flare} (\lambda_i)$ or $f_{\rm em} (\lambda_i)$ is:

\begin{equation} \label{equ:sig}
\sigma_i = \frac{1}{a}\sqrt{\sigma_{\rm flare}^2 (\lambda_i) + a^2\sigma_{{\rm Q},0}^2 (\lambda_i)}
\end{equation}
Here, $\sigma_{\rm flare} (\lambda_i) $ is the error of $f(\lambda_i)$ and $\sigma_{{\rm Q},0} (\lambda_i)$ is the error of $f_{{\rm Q},0} (\lambda_i)$, both of which are given by the LAMOST pipeline.

\section{Stellar Properties}
\subsection{The Radial Velocity ($RV$) and the Projected Rotational Velocity ($V_{\rm e}\sin i$)} \label{sec:vel}
In APOGEE-2 DR17 \citep{2022ApJS..259...35A}, the $T_{\rm eff} = 3645.9 \pm 5.5$ K, $\log g = 4.656 \pm 0.015$ dex, [Fe/H] $= 0.0269 \pm 0.0072$ dex, and [$\alpha$/Fe] = $-0.0510\pm0.0025$ dex.
We used the high-resolution PHOENIX spectral templates \citep{2013A&A...553A...6H} to calculate the radial velocity ($RV$) and the projected rotational velocity ($V_e\sin i$) of the star from its LAMOST MRS in quiescence.
The template liberary has a $T_{\rm eff}$ step of 100 K and a $\log g$ step of 0.5 dex. We selected eight templates with $T_{\rm eff} = 3500, 3600, 3700, 3800, 3900$ K and $\log g = 4.5, 5$ dex. Their [Fe/H] and [$\alpha$/Fe] were set to 0. 
\par
For each PHOENIX spectrum ($S_P$) we convolved it with the LAMOST instrumental profile ($f_P$) to obtain the LAMOST template $f_L$: $f_L = S_P \otimes f_P$. The profile of Th I $\lambda$6459 from the LAMOST arc lamp spectrum was used as the LAMOST instrumental profile, because it is between Ca I $\lambda$6441 and Ca I $\lambda$6464, which are the strongest lines in the spectrum and thus will be used to calculate  $V_{\rm e}\sin i$.
We matched the quiescence spectrum  $f_{{\rm Q},0} (\lambda)$ with the template spectrum $f_L$ with RV from -20 km\,s$^{-1}$ to 0 km\,s$^{-1}$ in a step of 0.5 km\,s$^{-1}$ by the cross-correlation method. All eight templates showed the $RV =-9.7 \pm 0.5$ km\,s$^{-1}$. 
\par
\begin{figure}[ht]
  \centering
   \includegraphics[scale=0.6]{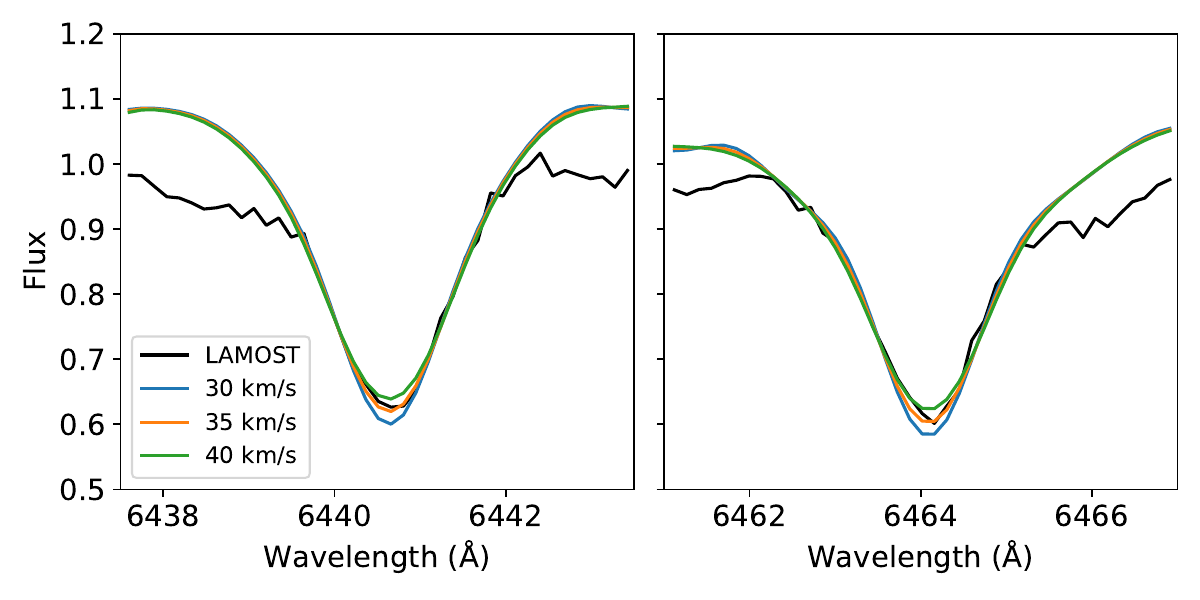}
  \caption{The rotational profiles of Ca I $\lambda$6441 and Ca I $\lambda$6464.  The black spectrum is the quiescence spectrum, while the blue, yellow and green curves are the PHOENIX spectrum of $T_{\rm eff} = 3700$ K and $\log g = 4.5$ dex, convolved  with $V_{\rm e}\sin i = 30, 35, 40$ km\,s$^{-1}$, respectively.
} 
\label{fig:rotv}
\end{figure}

To calculate the $V_{\rm e}\sin i$, the rotational broadening profile $f_{\rm rot}(V_{\rm e}\sin i)$ was from \citep{2005oasp.book.....G}, with the linear limb-darkening coefficient of $\epsilon= 0.8$, which was obtained from the linear limb-darkening coefficients of the $R$ band given in Table 40 of \citet{2000A&A...363.1081C}.
For each PHOENIX spectrum, the rotational templates were generated by $f_T(V_{\rm e}\sin i) = f_L \otimes f_{rot}(V_{\rm e}\sin i)$ with $V_{\rm e}\sin i$ from 0 km\,s$^{-1}$ to 60 km\,s$^{-1}$ in a step of 5 km\,s$^{-1}$. By comparing Ca I $\lambda$6441 and Ca I $\lambda$6464 profiles in the quiescence spectrum  $f_{{\rm Q},0} (\lambda)$ with these templates, the result $V_{\rm e}\sin i$ were obtained. The slowest  $V_{\rm e}\sin i = 30$ km\,s$^{-1}$ are obtained from templates with $T_{\rm eff} = 3500$ K, while the fastest $V_{\rm e}\sin i = 40$ km\,s$^{-1}$ are obtained from templates with $T_{\rm eff} = 3900$ K. Therefore, $V_{\rm e}\sin i = 35 \pm 5$ km\,s$^{-1}$ is adopted, which is similar to $V_{\rm e}\sin i = 31.1$ km\,s$^{-1}$ as given by the APOGEE-2 DR17. In Fig.~\ref{fig:rotv}, Ca I $\lambda$6441 and Ca I $\lambda$6464 profiles are shown. In either panel, the black spectrum is the quiescence spectrum, while the blue, yellow and green lines are the PHOENIX spectrum of $T_{\rm eff} = 3700$ K and $\log g = 4.5$ dex, convolved  with $V_{\rm e}\sin i = 30, 35, 40$ km\,s$^{-1}$, respectively.


\subsection{Stellar Rotational Period, Activity and Age}\label{sec:pdactage}
We obtained the TESS light curves using the \textsc{lightkurve} package \citep{2018ascl.soft12013L}. There are three light curves: two light curves at a 2-minute cadence from Sector 15 and 16 provided by the Science Processing Operations Center (SPOC; \citep{2016SPIE.9913E..3EJ}) and one light curve at a 30-minute cadence from Sector 22 provided by TESS-SPOC \citep{2020RNAAS...4..201C}. We detected 12, 15 and 18 flares in light curves from Sector 15, 16 and 22, respectively, using the method given by \citep{2023RAA....23a5016L}. The method firstly fitted the continuum of a light curve by a cubic B-spline, and after subtracting the fitted cubic B-spline, a stellar flare is detected by more than two continuous points higher than 5$\sigma$. The method tries to find big flares and then would overlook smaller ones.
\par
\begin{figure*}[ht]
  \centering
   \includegraphics[scale=0.6]{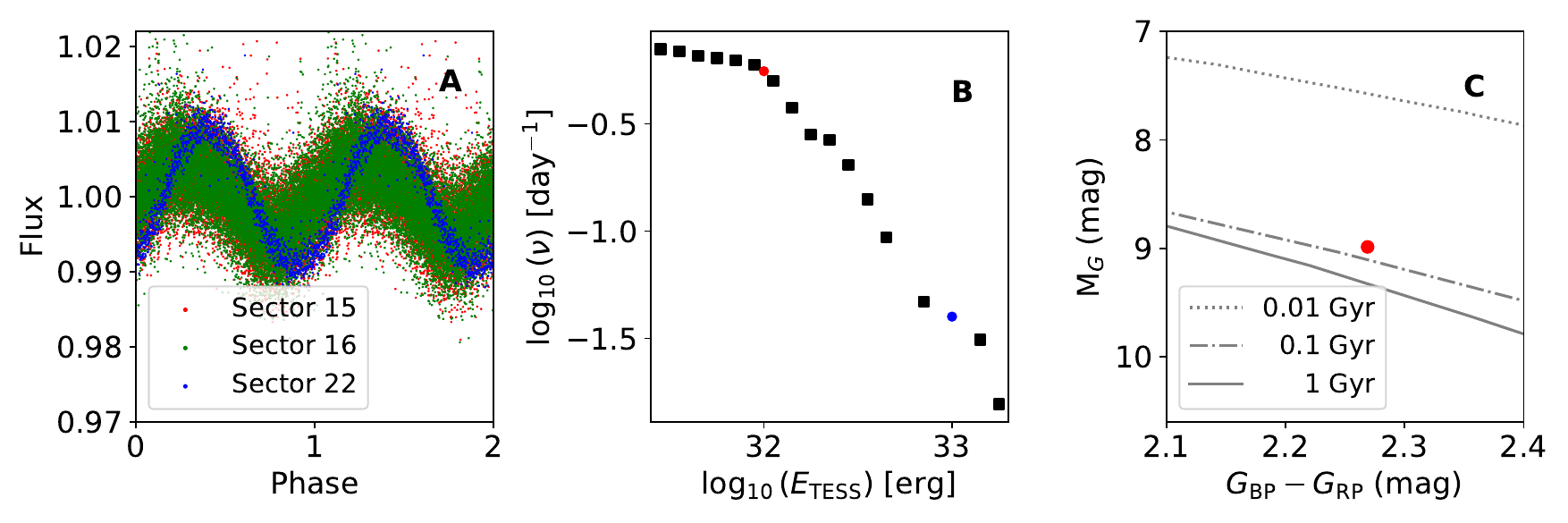}    
  \caption{J1332+5057 is a young and active star. (\textbf{A}) The TESS light curves folded by the period of 0.423051 days. Light curves from TESS Sector 15, 16 and 22 are shown in red, green and blue, respectively. (\textbf{B}) the FFD in the TESS band.
  The red circle is the flare frequency of $E_{\rm TESS} \geqslant 10^{32}$ erg, while the blue circle is the flare frequency of $E_{\rm TESS} \geqslant 10^{33}$ erg, which are once every 1.8 days and once every 25 days, respectively. 
   (\textbf{C}) The Gaia $G_{\rm BP} - G_{\rm RP}$ v.s. the absolute magnitude of Gaia $G$ ($M_G$) diagram. The red circle is J1332+5057 and the PARSEC isochrones of 0.01, 0.1 and 1 Gyr are shown by dotted, dash-dotted and solid lines, respectively. The error of $M_G$ is smaller than the circle size.} 
  \label{fig:pdffd}
\end{figure*}

The rotational periods were calculated from flare-removed light curves of Sector 15, 16 and 22 by  the Lomb-Scargle periodogram method (LS)\citep{1976Ap&SS..39..447L,1982ApJ...263..835S} using the python package \textsc{astropy.timeseries.LombScargle} \citep{astropy:2022}, and the result periods are 0.423080 days, 0.423005 days and 0.423069 days, respectively. Therefore, the adopted period is  0.423051$\pm$0.000033 days. The folded light curves from Sector 15, 16 and 22 are shown in the left panel of Fig.~\ref{fig:pdffd} in red, green and blue, respectively. The different light curve shapes between Sector 22 and Sector 15 and 16 may result from spot variations. 

In TESS light curves, flare energies were estimated from the equivalent duration (ED) \citep{1972Ap&SS..19...75G}, which is defined as: $ED = \int\frac{f(t)-f_0}{f_0}dt$, where $f(t)$ is the stellar flux during flare, $f_0$ is the stellar flux in quiescence, and $t$ is the time. Thus, the flare energy in the TESS band is $E_{\rm TESS} = ED \times f_0$. The stellar flux in quiescence is the median value of TESS light curves, the TESS spectral response curve and the passband zero are from \citep{2015ApJ...809...77S}, and the stellar distance is estimated from the parallax $\varpi$ in Gaia DR3, which is $1000/\varpi$ in pc. Finally, $E_{\rm TESS}$ can be calculated from the above qualities. 
\par
The stellar flare frequency distribution (FFD) \citep{1972Ap&SS..19...75G,2020AJ....159...60G} is often used to show the frequency of flares with the energy higher than a given threshold. The FFD of J1332+5057 derived from TESS flares is shown in the middle panel of Fig.~\ref{fig:pdffd}. We obtained the flare frequencies of $E_{\rm TESS} \geqslant 10^{32}$ erg and $E_{\rm TESS} \geqslant 10^{33}$ erg by interpolation, which are shown by the red and blue circles, respectively in the panel B of Fig. \ref{fig:pdffd}. A flare with $E_{\rm TESS}\geqslant10^{32}$ erg could occur once every 1.8 days, whereas a flare with $E_{\rm TESS}\geqslant10^{33}$ erg could occur once every 25 days.
\par
The stellar critical period of saturation is $P_{\rm sat} = 1.6 (L_{\star}/L_{\odot})^{-\frac{1}{2}}$ \citep{2014ApJ...794..144R}, which is $\sim 8.36$ days by using the $L_{\star}$ in Table~\ref{tab:pars}, much longer than $P_{\rm rot}$ ($\sim$0.423051 days). The X-ray flux in quiescence was obtained from \citep{2022A&A...664A.105F}, and $\log_{10}(L_X/L_{\star}) \sim -3.1$. 
As a result, both $P_{\rm rot}$ and the X-ray flux indicate that the stellar magnetic activity is saturated \citep{2003A&A...397..147P,2018MNRAS.479.2351W, 2014ApJ...794..144R}.
\par
The Gaia $G_{\rm BP} - G_{\rm RP}$ v.s. the absolute magnitude of Gaia $G$ ($M_G$) diagram is shown in the right panel of Fig.~\ref{fig:pdffd}. The red circle is J1332+5057 and the PARSEC isochrones \citep{2014MNRAS.444.2525C} of 0.01, 0.1 and 1 Gyr are shown by dotted, dash-dotted and solid lines, respectively. From this diagram we can see that the stellar age is less than 0.1 Gyr, making it very young.

\subsection{Stellar parameters} \label{sec:paras}
The stellar luminosity $L_{\star}$ was calculated by the following equations:

\begin{equation}\label{equ:Mg}
M_G = G + 5 \log_{10}(\varpi) - 10
\end{equation}
\begin{equation}\label{equ:lum}
M_{\star,\rm bol} = M_G + {\rm BC}_G 
\end{equation}
\begin{equation}\label{equ:lum_mg}
\log_{10}(L_{\star}/L_{\odot}) = 0.4(4.74 - M_{\star,\rm bol} )
\end{equation}

Here, $G$ in mag and parallax $\varpi$ in milliarcsecond are from Gaia DR3; ${\rm BC}_G = -0.656$ mag was obtained from \citep{2019A&A...632A.105C} with $T_{\star,\rm eff} = 3645.9$ K and  $\log g_{\star} = 4.656$.
\par
The following equations were used to calculate stellar parameters:

\begin{equation}\label{equ:rt}
L_{\star}/L_{\odot} =(R_{\star}/R_{\odot})^2 (T_{\star,\rm eff}/T_{\odot, \rm eff})^4
\end{equation}
\begin{equation}\label{equ:g}
g_{\star}/g_{\odot} = (M_{\star}/M_{\odot}) (R_{\star}/R_{\odot})^{-2}
\end{equation}
\begin{equation}\label{equ:ve0}
V_{\rm e} =  2\pi R_{\star}/P_{\rm rot}
\end{equation}
\begin{equation}\label{equ:i}
i = \arcsin(V_{\rm e} \sin i /V_{\rm e})
\end{equation}

Here, $L_{\star}$, $R_{\star}$, $T_{\star,\rm eff}$, and $g_{\star}$ are the stellar luminosity, radius, surface effective temperature, and gravity, respectively;  $L_{\odot}$, $R_{\odot}$, $T_{\odot, \rm eff}$, and $g_{\odot}$ are the solar luminosity, radius, surface effective temperature, and gravity, respectively, from \citep{2015arXiv151007674M};  $V_{\rm e}$, $i$, and $P_{\rm rot}$ are the stellar rotational velocity, inclination angle and rotational period, respectively. 
\par
From Equation~\ref{equ:rt} and~\ref{equ:g}, we can obtain:

\begin{eqnarray}
M_{\star}/M_{\odot} &= (L_{\star}/L_{\odot}) (g_{\star}/g_{\odot}) (T_{\star,\rm eff} / T_{\odot, \rm eff})^{-4} \\
R_{\star}/R_{\odot} &=  (L_{\star}/L_{\odot})^{\frac{1}{2}} (T_{\star, \rm eff} / T_{\odot, \rm eff})^{-2}\\
V_{\rm e} &=  (2\pi R_{\odot}/P_{\rm rot} ) (L_{\star}/L_{\odot})^{\frac{1}{2}} (T_{\star, \rm eff} / T_{\odot, \rm eff})^{-2} \label{equ:ve}
\end{eqnarray}

Therefore, the stellar mass ($M_{\star}$) and radius ($R_{\star}$) with errors were calculated from the stellar luminosity ($L_{\star}$), gravity ($g_{\star}$) and effective temperature ($T_{\star, \rm eff}$) with errors by \textsc{emcee} \citep{2013PASP..125..306F}. With the additional rotational period ($P_{\rm rot}$) with error, the rotational velocity ($V_{\rm e}$) with error can also be calculated. Finally, the inclination angle $i$ with error can be calculated from $V_{\rm e}$ and $V_{\rm e}\sin i $ with errors from Equation~\ref{equ:i}. The results are given in Table~\ref{tab:pars} and the posterior distributions of the stellar parameters are shown in Appendix Fig. \ref{fig:stellar_corn}.

\begin{table}
\centering
\caption{Stellar Parameters}
\label{tab:pars}%
\begin{tabular}{lll}\\
\hline
Parameter &Value & Source\\\hline
$\alpha$     & 203.055260$^{\circ}$ & \citep{gaia3}\\ 
$\delta$      & +50.950616$^{\circ}$ & \citep{gaia3}\\
$G_{\rm BP}$ & 14.337 mag               & \citep{gaia3}\\
$G_{\rm RP}$ & 12.068 mag               & \citep{gaia3}\\
$G$                 &13.153 mag               &\citep{gaia3}\\
$\varpi$           &14.6749 $\pm$ 0.0117 mas &\citep{gaia3}\\
SpT                 & dM2                                     & \citep{2015RAA....15.1095L} \\
$T_{\star,\rm eff}$   & 3645.9 $\pm$ 100 K               & \citep{2022ApJS..259...35A}\\
$\log g_{\star}$           & 4.656 $\pm$ 0.1 dex          & \citep{2022ApJS..259...35A}\\\hline
Age & $< 100$ Myr & This work \\
$P_{\rm rot}$   & 0.423051$\pm$ 0.000033 days     & This work \\
$RV$ & $-9.7\pm0.5$ km\,s$^{-1}$ & This work \\
$V_e \sin i$          & 35 $\pm$ 5 km\,s$^{-1}$                & This work\\
$L_{\star}$ & 0.036665 $\pm$ 0.000058 $L_{\odot}$ & This work \\
$M_{\star}$      & $0.38_{-0.085}^{+0.110}$ M$_{\odot}$ & This work \\
$R_{\star}$      &  $0.48 \pm 0.026$ R$_{\odot}$  & This work \\
$V_e$             & $57.4_{-2.9}^{+3.2}$ km\,s$^{-1}$              & This work \\
$i$                   & $37.4_{-6.3}^{+7.2}$ $^{\circ}$           & This work \\\hline
\end{tabular}
\tablecomments{The errors of $T_{\star,\rm eff}$ and $\log g_{\star}$ in \citet{2022ApJS..259...35A} are 5.5 K and 0.015 dex, respectively. The larger errors of 100 K and 0.1 dex for $T_{\star,\rm eff}$ and $\log g_{\star}$, respectively, are used in this work.}
\end{table}


\section{Radial Velocities of Mg I lines}
\begin{figure} 
	\centering
	\includegraphics[width=0.8\textwidth]{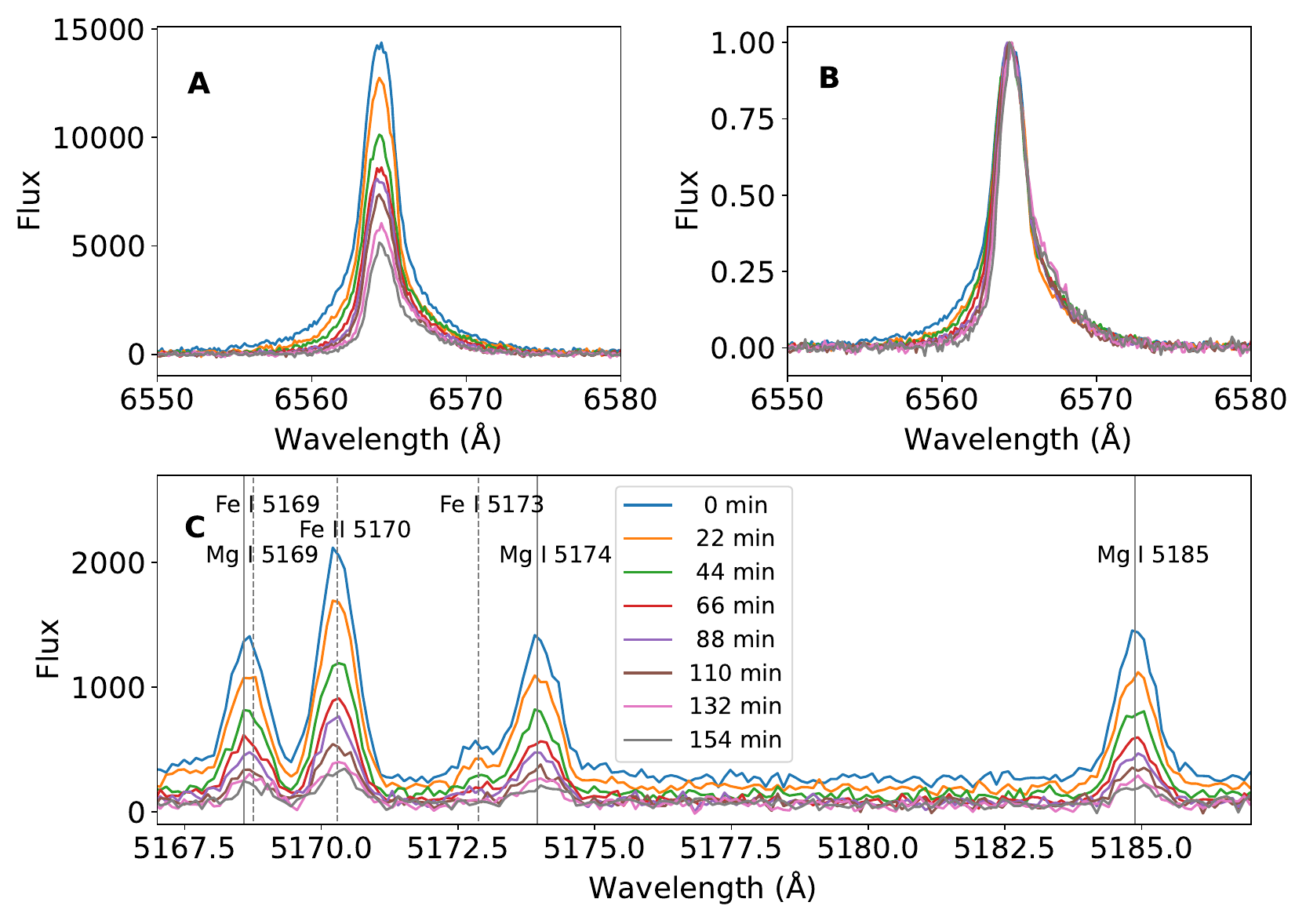} 

	\caption{H$\alpha$ and Mg I b emissions.
		(\textbf{A}) H$\alpha$ from eight individual exposures.  (\textbf{B}) Same as (\textbf{A}), but divided by their maximum fluxes. 
		(\textbf{C}) The Mg I b triplet at 5185, 5174 and 5169 ${\rm \AA}$, and Fe II $\lambda$5170 emission lines. The Mg I $\lambda$5169  may be contaminated by Fe I $\lambda$5169.
		The legend is shown in \textbf{C}.}
	\label{fig:HaMg} 
\end{figure}

We identified 62 emission lines in the first flare-only spectrum, and they are available online \footnote{\url{https://nadc.china-vo.org/res/r101659/}}. The chromospheric Mg I b triplet line and H$\alpha$ are both very strong and can be seen in all eight flare-only spectra (see Fig. \ref{fig:HaMg}). 
While the H$\alpha$ profile shows variations, the Mg I b profiles are stable and narrower than H$\alpha$. This suggests that the H$\alpha$ and Mg I b emission lines originate from different regions. In fact, on the Sun, the Mg I b emission line typically originate around the coolest chromospheric regions \citep{2017A&A...604A..50S}, whereas H$\alpha$ sources are more complex.
The Mg I $\lambda$5169 is contaminated by Fe I $\lambda$5169 and also very near to Fe II $\lambda$5170 (see Fig. \ref{fig:HaMg}C), so the Mg I $\lambda$5174 and Mg I $\lambda$5185 were adopted to locate the flare. 

We assume that either profile of Mg I $\lambda$5174 and  $\lambda$5185 is a Gaussian function, and their central  radial velocities are same.
\begin{eqnarray}\label{equ:mg}
P_1(v_1) &= a_1 \exp(-\frac{(v_1- v_c)^2}{2\sigma_1^2}) + b_1\\
P_2(v_2) &= a_2 \exp(-\frac{(v_2-v_c)^2}{2\sigma_2^2}) +b_2
\end{eqnarray}

where, $P_1(v_1)$ and $P_2(v_2)$ are Gaussian functions for Mg I $\lambda$5174 and  $\lambda$5185, respectively;  $v_c$ is the radial velocity;  the $a_1$, $a_2$, $\sigma_1$, $\sigma_2$, $b_1$ and $b_2$ are constants to be fitted.
The routine \textsc{scipy.optimize.curve\_fit} \citep{2020NatMe..17..261V} was used to obtain the parameters and their errors. The fitted radial velocities $v_c$ and errors are given in Table~\ref{tab:flr} and shown in Fig.~\ref{fig:mg}. During the observation, the systematic error of radial velocities from LAMOST MRS is about 0.18 km\,s$^{-1}$ (see Fig. \ref{fig:drv} in APPENDIX \ref{sec:lamost_rv}), so the fitted radial velocities with errors should be from Mg I emission lines.

We required that the phase $\omega$ of the first LAMOST spectrum is 0, and at $\omega=0$, the stellar longitude that crosses the line of sight to the stellar center is $\phi=0^{\circ}$. We assume that the flare location on the stellar surface is ($\phi_0$, $\theta_0$). In the stellar coordinate system, the direction of the line of sight is $(-\sin i, 0, -\cos i)$, and the rotational velocity of the flare location at the time $t$ is: $(-V_{\rm e} \cos \theta_0 \sin(\phi_0+2\pi\frac{t}{P_{\rm rot}}),  V_{\rm e} \cos \theta_0 \cos(\phi_0+2\pi\frac{t}{P_{\rm rot}}), 0)$. Then the projected rotational velocity of the flare location on the line of sight is:

\begin{eqnarray} 
v_1(t) &=& (-V_{\rm e} \cos \theta_0 \sin(\phi_0+2\pi\frac{t}{P_{\rm rot}}), V_{\rm e} \cos \theta_0 \cos(\phi_0+2\pi\frac{t}{P_{\rm rot}}), 0) \bullet (-\sin i, 0, -\cos i) \nonumber\\
      &=& V_{\rm e}\sin i \cdot \cos \theta_0\cdot\sin(\phi_0+2\pi\frac{t}{P_{\rm rot}}) \label{equ:vr}
\end{eqnarray}
Here, $\bullet$ is the dot product operator. 
\par

We assume that in the direction vertical to the stellar surface, the cool chromospheric materials have a  net motion, $v_{\bot}$. In the stellar coordinate system, the direction of $v_{\bot}$ is: $(\cos \theta_0 \cos (\phi_0+2\pi\frac{t}{P_{\rm rot}}),   \cos \theta_0 \sin (\phi_0+2\pi\frac{t}{P_{\rm rot}}), \sin \theta_0)$. Therefore, the projection of $\overrightarrow{v_{\bot}}$ on the line of sight is:

\begin{eqnarray}
v_2(t) &= & v_{\bot} \cdot  (\cos \theta_0 \cos (\phi_0+2\pi\frac{t}{P_{\rm rot}}),   \cos \theta_0 \sin (\phi_0+2\pi\frac{t}{P_{\rm rot}}), \sin \theta_0) \bullet (-\sin i, 0, -\cos i)  \nonumber\\
           &=& -v_{\bot} \cdot \left[  \sin i \cos \theta_0 \cos (\phi_0+2\pi\frac{t}{P_{\rm rot}}) +   \sin \theta_0  \cos i \right] \label{equ:vv}
\end{eqnarray}

As a result, the radial velocity $v(t)$ of Mg I emission lines is: 

\begin{equation} \label{equ:vl}
v(t) = v_1(t)+v_2(t)+RV.
\end{equation}

In Equation \ref{equ:vl}, $v(t)$ consists of a time-varying component and a constant component. As the stellar inclination $i$ is fixed, Equation \ref{equ:vr} and \ref{equ:vv} show that the amplitude of the time-varying component of  $v(t)$ is determined by $\cos \theta_0$. This implies that the lower the latitude of the flare location, the greater the variation amplitude. The shape of the $v(t)$ follows a trigonometric function modulated by the time-dependent longitude $\phi_0+2\pi\frac{t}{P_{\rm rot}}$. The constant component of $v(t)$ is primarily determined by $RV$ and $v_{\bot}  \sin \theta_0 $ in Equation \ref{equ:vv}. The higher the latitude of the flare location, the larger the $v_{\bot}$,  and the greater the influence on the constant component of $v(t)$.

\section{Flare Location}\label{sec:acu}
The Markov Chain Monte Carlo (MCMC) method was used to obtain the parameters of Equation~\ref{equ:vl} by the routine \textsc{emcee}. 
The log-likelihood function is 

\begin{equation}\label{equ:llf}
\ln L =-\frac{1}{2}\displaystyle \sum_{i=1}^8 \bigg [ \frac{v(t_i) -v_i}{\sigma_i} \bigg ]^2
\end{equation}

Here, $t_i$ is the observation time of the $i$th LAMOST spectrum, $v(t_i)$ is the $v(t)$ in Equation~\ref{equ:vl} at $t_i$, $v_i$ is the observational radial velocity of Mg I lines at $t_i$ in Table~\ref{tab:flr}, and $\sigma_i$ is the error of $v_i$. 
We assume the $T_{\rm eff}$, $\varpi$, $P_{\rm rot}$, $RV$ and $V_{\rm e}\sin i$ in Table~\ref{tab:pars} follow normal distributions, and the sum of logarithms of their probability density functions are used as the logarithm of the prior distribution, which is denoted as $\ln L_{\rm pri}$. Therefore, the logarithm of the posterior distribution is:

\begin{equation}\label{equ:pos}
\ln L_{\rm pos} = \ln L + \ln L_{\rm pri}
\end{equation}
The number of walkers was set to 32, and walked 40000 steps.
First, we ran the MCMC algorithm, and calculated the medians of parameter samples. We then used these medians, to which we added a small random perturbation, as the initial input and ran the MCMC algorithm again. The autocorrelation time of the obtained parameter samples were between 110 and 190 steps. Therefore, for each  parameter sample, we discarded the first 380 sample points ($190 \times 2 = 380$) and then selected one sample point every 55 points ($110 \div 2 = 55$). The parameters of the posterior distributions are shown in Fig. \ref{fig:corn} in  Appendix~\ref{sec:pdf}, and their median values and errors are given respectively by the 50th and 16th and 84th percentiles of the parameter samples.
\begin{figure}
  \centering
   \includegraphics[scale=0.5]{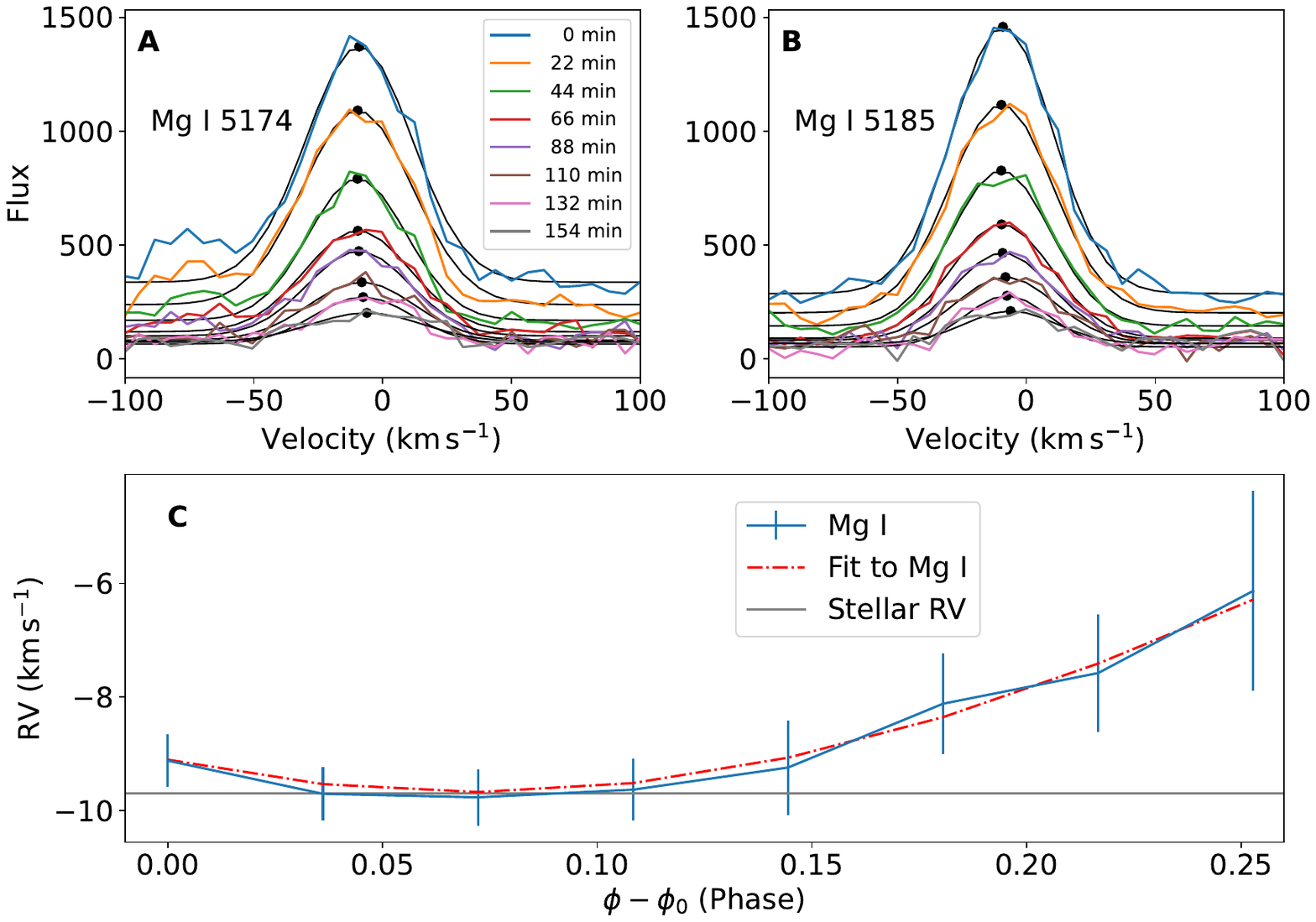}
  \caption{The radial velocities obtained from Mg I $\lambda$5174 and Mg I $\lambda$5185. The Mg I $\lambda$5174 and Mg I $\lambda$5185 emission lines from eight flare-only spectra are shown in (\textbf{A}) and (\textbf{B}), respectively, and the fitted profiles are shown by thin black lines. The fitted Gaussian peaks are shown by black circles. The bump on the left wing of Mg I $\lambda$5174 is the Fe I $\lambda$5173 emission line. (\textbf{C}) The radial velocities of Mg I, the stellar radial velocity (RV) and the fitted radial velocities by Equation~\ref{equ:vl} are shown by the blue, grey and red dash-dotted lines, respectively.} 
  \label{fig:mg}
\end{figure}
\begin{figure}
  \centering
   \includegraphics[scale=0.7]{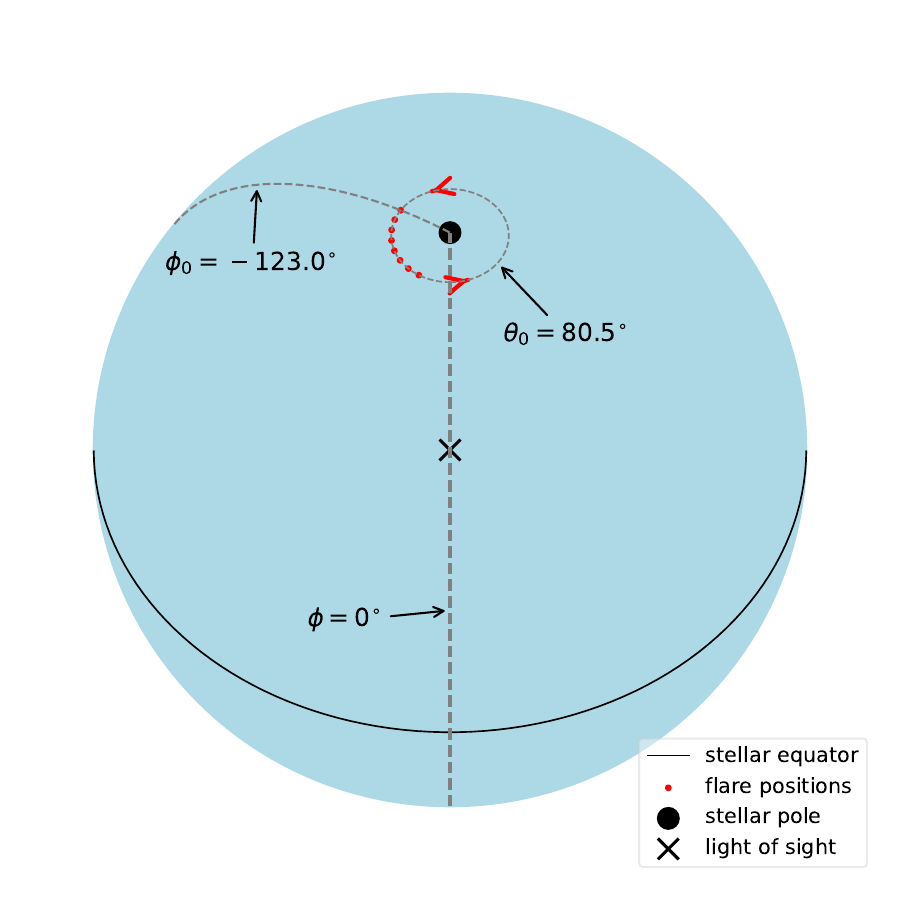}
  \caption{The stellar surface projection of J1332+5057 along the line of sight at the phase 0. The flare locations of eight phases  obtained from Equation~\ref{equ:pos} are shown by red circles. The latitude $\theta_0 = 80.5^{\circ}$ and the longitude $\phi_0 = -123.0^{\circ}$ and $\phi = 0^{\circ}$ at the phase 0 are shown by gray dashed lines. The stellar rotation direction is shown by two red arrows and the line of sight is shown by the black cross.} 
  \label{fig:flarepos}
\end{figure}

The fitted latitude is $\theta_0 = 80.5^{+2.9}_{-3.2} $ in degree, the fitted longitude is $\phi_0 = -123.0^{+8.0}_{-5.8} $ in degree, and the fitted vertical velocity of material is $v_{\bot} = -7.5\pm2.3$ km\,s$^{-1}$. In Fig.~\ref{fig:flarepos}, flare location $(-123.0^{\circ}, 80.5^{\circ})$ of eight phases are shown by red circles and the fitted $v(t)$ is shown by the red dash-dotted line in Fig.~\ref{fig:mg}.

The flare latitude $\theta_0$ can also be estimated (see Appendix \ref{sec:est}) by:
\begin{equation} \label{equ:est}
\theta_0 \geqslant \arccos( \frac{\bigtriangleup v}{V_{\rm e}\sin i \times [1-\cos(2\pi \bigtriangleup\omega)]})
\end{equation}
where, $\bigtriangleup v$ and $\bigtriangleup\omega$ are the differences of radial velocities and phases of two different times, respectively, and the two radial velocities should be in the same monotonic interval. From Table~\ref{tab:flr}, the maximum $\bigtriangleup v \approx |-6.1 - (-9.7)| = 3.6$ km\,s$^{-1}$, and their time dfference is $154-44 =110$ minutes. Using Equation~\ref{equ:est}, we obtain $\theta_0 \gtrsim 79.7^{\circ}$, which is well consistent with the result $\theta_0=80.5^{+3.0}_{-3.2}$ in degree given by rigorous calculations. In fact, Equation ~\ref{equ:est} implies that the smaller the $\bigtriangleup v$, the higher the latitude $\theta_0$.

\begin{table*}
  \centering
  \caption{\textbf{radial velocities (RV) of the Mg I b emission line.}}
	\label{tab:flr} 

\begin{tabular}{|l|l|l|l|l|l|l|l|l|}
\hline
Time (min)  &  0    &    22    &    44    &    66    &    88    &    110    &    132    &    154   \\\hline
RV (km\,s$^{-1}$) & -9.1$\pm$0.5 &-9.6$\pm$0.5 &-9.7$\pm$0.5 & -9.6$\pm$0.5 &-9.2$\pm$0.8 & -8.1$\pm$0.9 & -7.6$\pm$1.0 &-6.1$\pm$1.8 \\\hline
\end{tabular}
\end{table*}


\section*{Discussion and Conclusion} \label{sec:con}

By establishing the Mg I b emission line as a practical tool for flare localization, the flare location on J1332+5057 is pinpointed at  $(-123.0^{+8.0}_{-5.8}, 80.5^{+2.9}_{-3.2})$ in degree, which is significantly different from solar flare locations.
Fast rotators tend to produce frequent and energetic stellar flares \citep{2019ApJS..241...29Y,2022ApJ...935..104M}, and their large spots tend to appear at high latitudes and even in polar regions \citep{2009A&ARv..17..251S,1992A&A...264L..13S}. If these flares originate from spots in high latitudes, their impacts on habitable planets would be significantly reduced. Indeed, some observational evidences suggest that stellar flares can occur at high latitudes \citep{2024A&A...682A.176B, 2021MNRAS.507.1723I, 2021NatAs...5..697V}. Therefore,  determining flare locations is crucial for 
advancing our understanding of the stellar magnetic activity, the assessment of exoplanet habitability and the stellar angular momentum loss. This work provides a method to locate flares.
\par
The data used in this work and supplementary materials are available at \url{https://nadc.china-vo.org/res/r101659/}.

\begin{acknowledgments}
The authors thank the anonymous referee very much for the valuable report that inspired us to improve this work. 
This work is supported by the National Natural Science Foundation of China (NSFC) with grant No. 12494571. 
G. P.~Zhou acknowledges NSFC with grant No. 12533010.
G. W.~Li acknowledges the support from the Chinese Space Station Telescope (CSST) project.
W.H.~Wang acknowledge the support from National Key Research and Development Program of China with grant No. 2024YFA1611501.

This work made use of the data from LAMOST (Large Sky Area Multi-Object Fiber Spectroscopic Telescope, also known as the Guoshoujing Telescope) (\url{https://cstr.cn/31118.02.LAMOST}). LAMOST is a Chinese national mega-science facility, operated by National Astronomical Observatories, Chinese Academy of Sciences.
\end{acknowledgments}

%

\vspace{5mm}
\facilities{LAMOST}


\software{astropy \citep{2013A&A...558A..33A,2018AJ....156..123A},  
          Lightkurve \citep{2018ascl.soft12013L}, 
          numpy \citep{harris2020array}, matplotlib \citep{Hunter:2007}, emcee \citep{2013PASP..125..306F}, and scipy \citep{2020NatMe..17..261V},
          corner \citep{corner16}
          }



\appendix
\section{Precision of Radial Velocity}\label{sec:lamost_rv}
\begin{figure}
  \centering
   \includegraphics[scale=1]{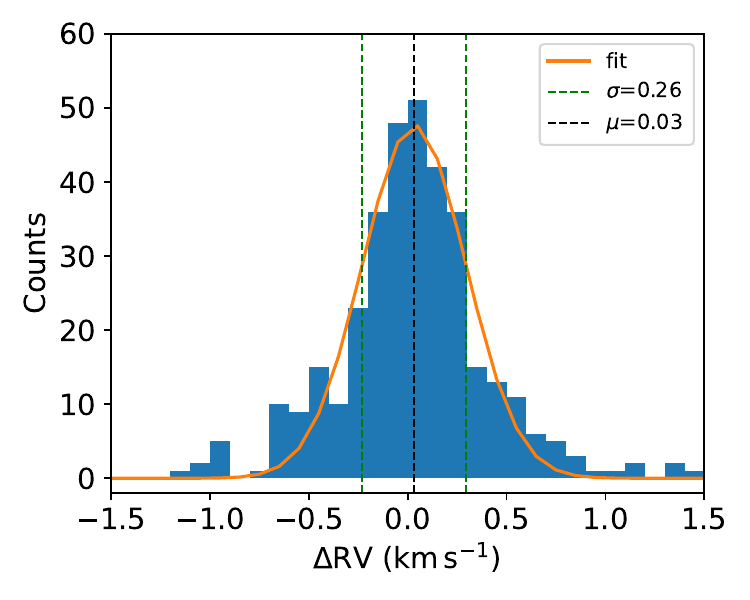}
  \caption{The distribution of radial velocity differences between LAMOST MRS. } 
  \label{fig:drv}
\end{figure}

The flare spectra of the star J1332+5057 were recorded by Spectrograph 3 of LAMOST on April 19, 2022. The LAMOST pipeline provides radial velocities labelled \textsf{rv\_lasp} for all MRS of individual exposures. We examined the radial velocities in \textsf{rv\_lasp} of all the stars that were observed simultaneously with J1332+5057 in the Spectrograph 3 of LAMOST. We found that a total of 50 stars had radial velocity measurements from all eight individual exposures. For each star, we calculated the velocity differences by subtracting the velocity of the first exposure from those of the other seven exposures. Fig. \ref{fig:drv} shows the distribution of these velocity differences for all 50 stars. We fitted this distribution with a Gaussian function, yielding a mean velocity of 0.03 km\,s$^{-1}$ and a standard deviation of 0.26 km\,s$^{-1}$. Therefore, the velocity dispersion for Spectrograph 3 was $0.26 / \sqrt{2} =$0.18 km\,s$^{-1}$ throughout the entire flaring duration from the star J1332+5057.

\section{Estimating Flare Latitude}\label{sec:est}
To solve the Equation~\ref{equ:vl}, too many stellar parameters are needed. Thus, we will give a concise formula to estimate the flare latitude from two observations. In Equation~\ref{equ:vl}, 

\begin{eqnarray}
  v(t) &= &v_1(t)+v_2(t)+RV\\
        & =& V_{\rm e}\sin i  \cos \theta_0 \sin(\phi_0+2\pi\frac{t}{P_{\rm rot}}) -v_{\bot} \left[ \cos \theta_0 \sin i \cos (\phi_0+2\pi\frac{t}{P_{\rm rot}}) +   \sin \theta_0  \cos i \right] +RV\\
        & =& \sin i \cos \theta_0 \left[V_{\rm e} \sin(\phi_0+2\pi\frac{t}{P_{\rm rot}}) -v_{\bot} \cos (\phi_0+2\pi\frac{t}{P_{\rm rot}})\right]  -  v_{\bot} \sin \theta_0  \cos i   + RV \\
        & =& \sin i \cos \theta_0 \sqrt{V_{\rm e}^2+v_{\bot} ^2}\sin(\phi_0+2\pi\frac{t}{P_{\rm rot}} +\phi')  -  v_{\bot} \sin \theta_0  \cos i   + RV\\
        & =&\sqrt{V_{\rm e}^2+v_{\bot} ^2} \sin i \cos \theta_0\sin(\phi_0+\phi'+2\pi\frac{t}{P_{\rm rot}})  -  v_{\bot} \sin \theta_0  \cos i   + RV
\end{eqnarray}
Here, $\tan \phi' = - \frac{ v_{\bot}}{V_{\rm e}}$.
\par
The radial velocities of Mg I emission lines from $t_1$ and $t_2$ are $v(t_1)$ and $v(t_2)$, respectively. Then

\begin{eqnarray}
v(t_2) - v(t_1) &=&  \sqrt{V_{\rm e}^2+v_{\bot} ^2} \sin i \cos \theta_0 \sin(\phi_0+\phi'+2\pi\frac{t_2}{P_{\rm rot}}) \\
                       &&\quad- \sqrt{V_{\rm e}^2+v_{\bot} ^2} \sin i \cos \theta_0\sin(\phi_0+\phi'+2\pi\frac{t_1}{P_{\rm rot}})  \\
                       &=&  \sqrt{V_{\rm e}^2+v_{\bot} ^2} \sin i \cos \theta_0 \left[\sin(\phi_0+\phi'+2\pi\frac{t_2}{P_{\rm rot}}) - \sin(\phi_0+\phi'+2\pi\frac{t_1}{P_{\rm rot}}) \right] \\
                       &= & \sqrt{1+(\frac{v_{\bot}}{V_{\rm e}})^2} (V_{\rm e}\sin i) \cos \theta_0 \left[\sin(\phi_0+\phi'+2\pi\frac{t_2}{P_{\rm rot}}) - \sin(\phi_0+\phi'+2\pi\frac{t_1}{P_{\rm rot}}) \right] 
\end{eqnarray}
That is,

\begin{equation}\label{equ:vdif}
\cos \theta_0 = \frac{v(t_2) - v(t_1)}{\sqrt{1+(\frac{v_{\bot}}{V_{\rm e}})^2} (V_{\rm e}\sin i)  \left[\sin(\phi_0+\phi'+2\pi\frac{t_2}{P_{\rm rot}}) - \sin(\phi_0+\phi'+2\pi\frac{t_1}{P_{\rm rot}}) \right]}
\end{equation}

Now,       

\begin{eqnarray}
&&(\sin(x+\triangle x) - \sin x ) - (1-\cos(\triangle x) ) \\
&=&    2\sin\frac{\triangle x}{2}\cos(x+\frac{\triangle x}{2}) - 2\sin^2\frac{\triangle x}{2}\\
    &=&  2\sin\frac{\triangle x}{2} \left[\cos(x+\frac{\triangle x}{2})  - \sin\frac{\triangle x}{2}\right]\\
    &= & 2\sin\frac{\triangle x}{2} \left[\sin(\frac{\pi}{2} - x- \frac{\triangle x}{2})  - \sin\frac{\triangle x}{2}\right]\\
    &= & 2\sin\frac{\triangle x}{2} \left[2\sin(\frac{\frac{\pi}{2} - x- \triangle x}{2}) \cos\frac{\frac{\pi}{2}- x}{2}\right] \label{equ:sucmul}
\end{eqnarray}

In the monotonic increasing interval $-\frac{\pi}{2} \leqslant x \leqslant x+\triangle x \leqslant \frac{\pi}{2}$ of the function $\sin x$, 
the $\frac{\triangle x}{2}$, $\frac{\frac{\pi}{2} - x- \triangle x}{2}$, and $\frac{ \frac{\pi}{2}- x}{2} $ are all in the interval $\left[0,  \frac{\pi}{2} \right]$, which mean that Equation~\ref{equ:sucmul} $\geqslant 0$. That is, for $-\frac{\pi}{2} \leqslant x \leqslant x+\triangle x \leqslant \frac{\pi}{2}$,

\begin{equation}\label{equ:int1}
\sin(x+\triangle x) - \sin x  \geqslant 1-\cos(\triangle x)
\end{equation}
Likewise, in the monotonic descreasing interval $\frac{\pi}{2} \leqslant x \leqslant x+\triangle x \leqslant \frac{3\pi}{2}$ of the function $\sin x$, 

\begin{equation}\label{equ:int2}
\sin x - \sin(x+\triangle x)   \geqslant 1-\cos(\triangle x)
\end{equation}

From Equation~\ref{equ:int1} and ~\ref{equ:int2}, we can see that for $x$ and $x+\triangle x$ in a same monotonic interval of the function $\sin x$, 

\begin{equation}\label{equ:mono}
 |\sin(x+\triangle x) -\sin x |  \geqslant 1-\cos(\triangle x)
\end{equation}
Therefore, for $v(t_1)$ and $v(t_2)$ in the same monotonic interval, Equation~\ref{equ:vdif} can be written as:

\begin{eqnarray}\label{equ:vdif1}
\cos \theta_0 &=& \frac{|v(t_2) - v(t_1)|}{\sqrt{1+(\frac{v_{\bot}}{V_{\rm e}})^2} (V_{\rm e}\sin i)  |\sin(\phi_0+\phi'+2\pi\frac{t_2}{P_{\rm rot}}) - \sin(\phi_0+\phi'+2\pi\frac{t_1}{P_{\rm rot}})|}\\
    &\leqslant& \frac{|v(t_2) - v(t_1)|}{\sqrt{1+(\frac{v_{\bot}}{V_{\rm e}})^2} (V_{\rm e}\sin i)  \left[1- \cos (2\pi\frac{t_2-t_1}{P_{\rm rot}})\right] }\\
    &\leqslant& \frac{|v(t_2) - v(t_1)|}{V_{\rm e}\sin i \times [1-\cos(2\pi\frac{t_2-t_1}{P_{\rm rot}})]} \\
                       &=& \frac{\bigtriangleup v}{V_{\rm e}\sin i \times [1-\cos(2\pi \bigtriangleup\omega)]}
\end{eqnarray}

where, $\bigtriangleup v = |v(t_2) - v(t_1)|$ and $\bigtriangleup\omega = \frac{|t_2-t_1|}{P_{\rm rot}}$.
As a result,

\begin{equation}
\theta_0 \geqslant \arccos( \frac{\bigtriangleup v}{V_{\rm e}\sin i \times [1-\cos(2\pi \bigtriangleup\omega)]})
\end{equation}
The equation shows that the flare latitude can be estimated from three observational quantities: $V_{\rm e}\sin i$, $\bigtriangleup v$ and $\bigtriangleup\omega$, and does not need to know the exactly $i$, $v_{\bot}$ and $\phi_0$.

\section{Posterior Distributions of the Stellar Parameters and Flare Location }\label{sec:pdf}
\begin{figure}
  \centering
   \includegraphics[scale=0.35]{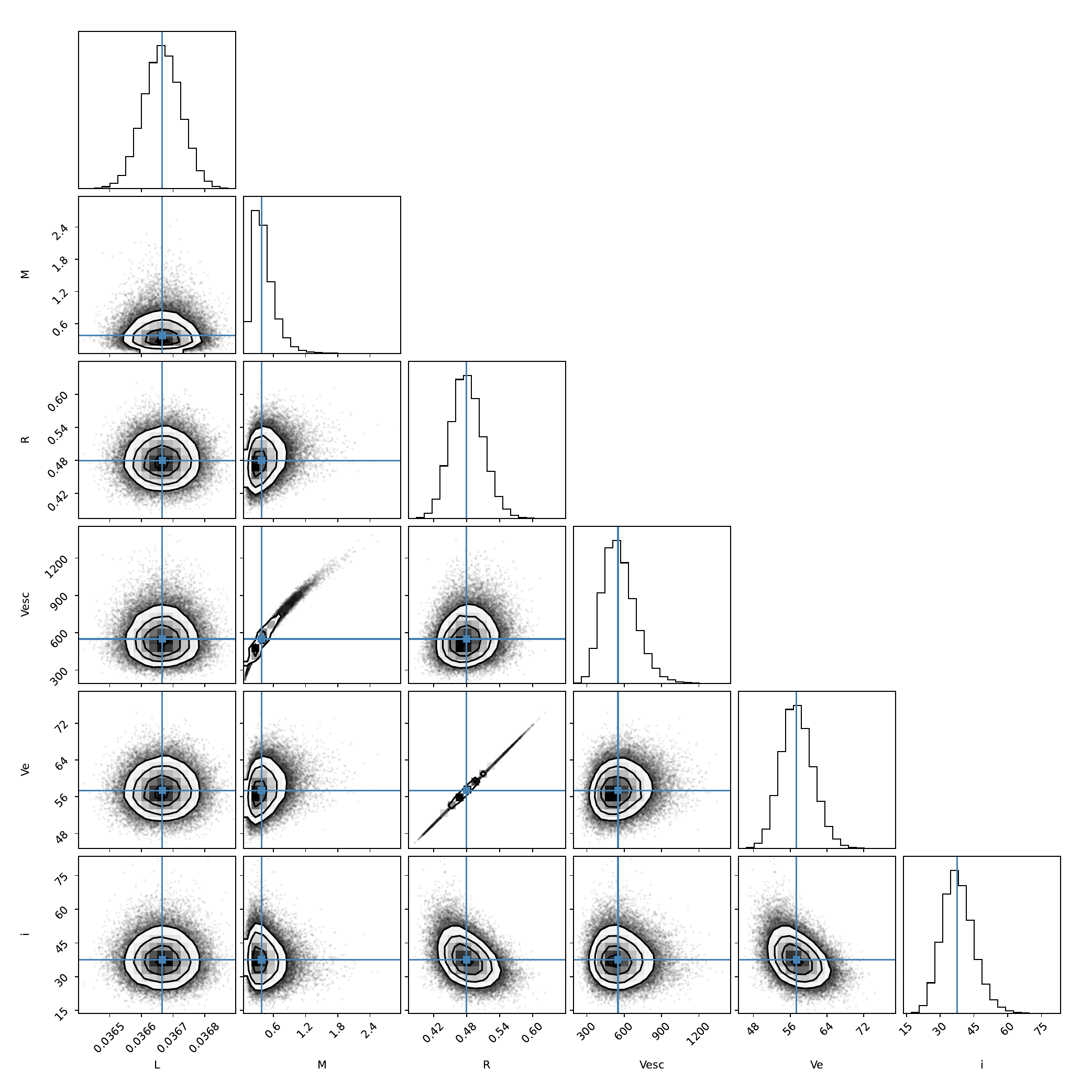}
  \caption{The posterior distributions of stellar parameters. Produced by the \textsl{corner} package \citep{corner16}.} 
  \label{fig:stellar_corn}
\end{figure}

The distributions of the stellar luminosity ($L_{\star}$), mass ($M_{\star}$),
radius ($R_{\star}$), the stellar rotational velocity at the equator ($V_{\rm e}$) and
the inclination angle ($i$) were calculated from the distributions
of the surface effective temperature $T_{\star, \rm eff}$, gravity $\log g_{\star}$, parallax $\varpi$,
rotational period $P$, and projected rotational speed $V_{\rm e} \sin i$ by
Equations \ref{equ:Mg} - \ref{equ:ve}. The stellar surface escape velocity $V_{\rm esc}$ was
also calculated. The results are shown in Fig. \ref{fig:stellar_corn}.

\begin{figure}
  \centering
   \includegraphics[scale=0.35]{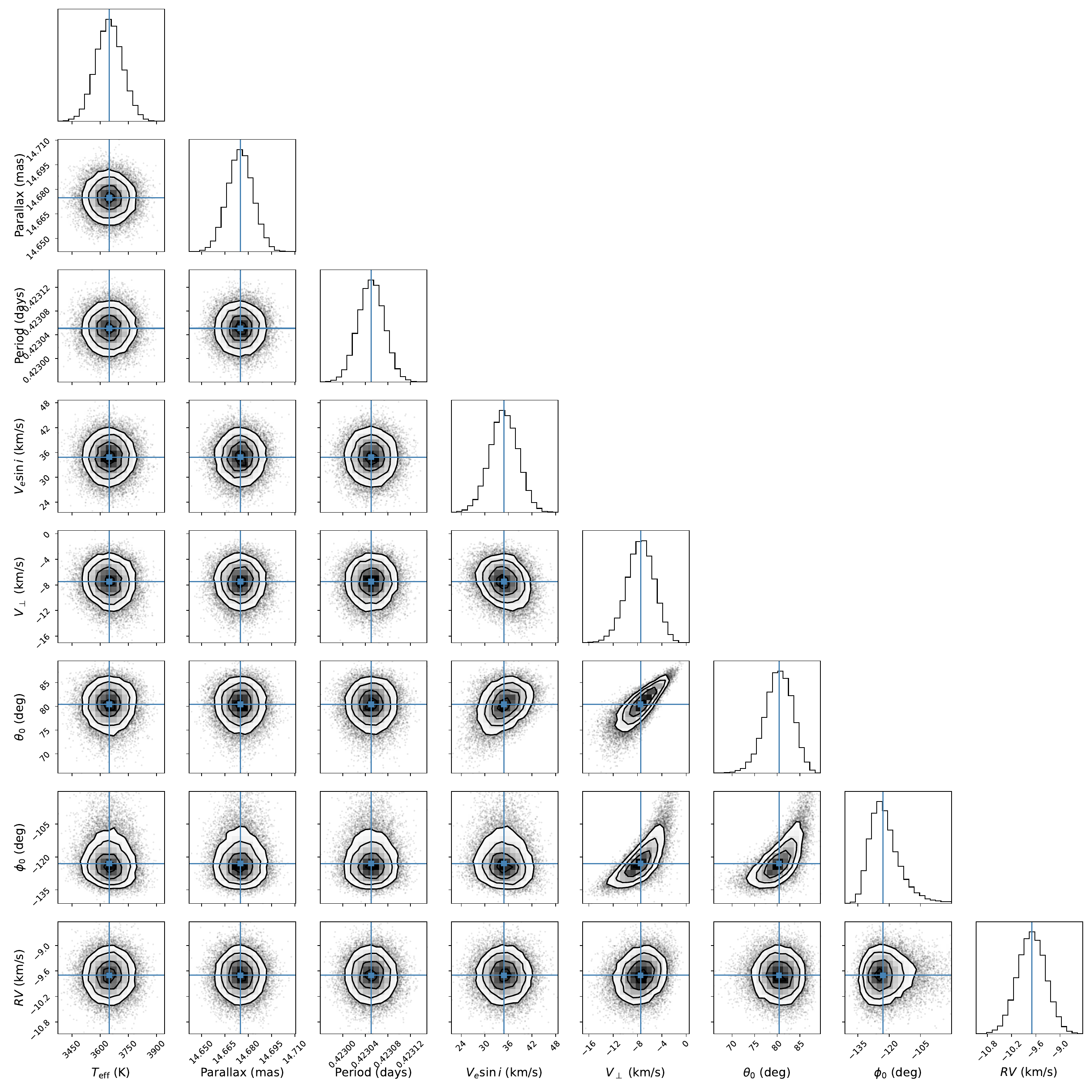}
  \caption{The posterior distributions of stellar parameters and the flare location. Produced by the \textsl{corner} package \citep{corner16}.} 
  \label{fig:corn}
\end{figure}

The parameters of the posterior distributions obtained from
Equation \ref{equ:pos} are shown in Fig. \ref{fig:corn}. The flare location is
specified.

\bibliography{polarflare}{}
\bibliographystyle{aasjournal}


 \end{CJK*}
\end{document}